\documentclass{article}

\usepackage[final]{neurips_2022_ml4ps}

\usepackage{mathtools}

\usepackage[utf8]{inputenc} 
\usepackage[T1]{fontenc}    
\usepackage{hyperref}       
\usepackage{url}            
\usepackage{booktabs}       
\usepackage{amsfonts}       
\usepackage{nicefrac}       
\usepackage{microtype}      
\usepackage{xcolor}         
\usepackage[T1]{fontenc}    
\usepackage{hyperref}       
\usepackage{url} 
\usepackage{booktabs}       
\usepackage{amsfonts}       
\usepackage{nicefrac}       
\usepackage{microtype}      
\usepackage{lipsum}		
\usepackage{graphicx}
\usepackage{natbib}
\usepackage{doi}
\usepackage{amsmath}
\usepackage{graphicx}
\usepackage{caption}
\usepackage{subcaption}
\usepackage{multirow}
\usepackage{float}
\usepackage{caption}
\usepackage{subcaption}

\setcitestyle{numbers,square}

\title{Point Cloud Generation using Transformer Encoders and Normalising Flows}

\author{
Benno Käch\\
Deutsches Elektronen-Synchrotron DESY\\  Germany\\
\texttt{benno.kaech@desy.de} \\
\AND
Dirk Krücker\\
Deutsches Elektronen-Synchrotron DESY\\ Germany\\
\texttt{dirk.kruecker@desy.de} \\
\And
Isabell-A. Melzer-Pellmann \\
Deutsches Elektronen-Synchrotron DESY\\  Germany\\
\texttt{isabell.melzer@desy.de} \\
}

\begin{document}
\maketitle

\begin{abstract}
    Data generation based on Machine Learning has become a major research topic in particle physics. This is due to the current Monte Carlo simulation approach being computationally challenging for future colliders, which will have a significantly higher luminosity. The generation of collider data is similar to point cloud generation, but arguably more difficult as there are complex correlations between the points which need to be modelled correctly. A refinement model consisting of normalising flows and transformer encoders is presented. The normalising flow output is corrected by a transformer encoder, which is adversarially trained against another transformer encoder discriminator/critic. The model reaches state-of-the-art performance while yielding a stable training.
    

\end{abstract}

\section{Introduction}
High Energy Physics (HEP) has benefited from the advances in Machine Learning (ML) since the analysis of HEP data is a high-dimensional multivariate problem. There have been multiple applications of ML to HEP~\cite{mlinhep}, although most of them were in the supervised approach to ML. In HEP, detailed simulations of the physical processes, which almost perfectly describe the details of the experimental measurement, are commonly available. These Monte Carlo simulations (MC) provide labeled data and are needed in large numbers to cover the extreme areas of the physical phase space. The simulations for the CMS detector at the Large Hadron Collider, for example, require about $50 \%$~\cite{mccms} of the current CMS computing budget. An even larger number of simulations will be needed for the coming high-luminosity phase of the LHC~\cite{hllhc}. Therefore, generative modelling with Deep Learning (DL) has sparked great interest in the HEP community.


In this study, the generation of jets is investigated using Normalising Flows (NFs)~\cite{tabak,rezende,kobyzev_2021,papama19} and an adversarially trained refinement setup consisting of two transformer~\cite{attention,transformers} encoder networks. The Jets in the dataset are described as point clouds in momentum space $\{(\eta^{(i)},\phi^{(i)},p_T^{(i)})_{i\leq 30}\}$, which motivates the use of transformers as attention is permutation invariant. 
One of the transformer encoders acts as a refinement model to enhance the NF output. The use of NFs is motivated by their stable convergence thanks to Maximum Likelihood training, however NFs often struggle to model all correlations present in the training data correctly and are not particularly well suited for point clouds. The use of the refinement network enhances the performance in the modeling of the correlations significantly.

This extended abstract first gives an overview over the models used, then the different training paradigms are briefly discussed. Finally, the results for different models are presented and compared.
\section{Dataset}
In this first study, the \emph{JetNet}~\cite{mpgan} top quark dataset~\cite{topquark} is used. We restrict ourselves to top quark jets due to their more complex sub-structure. The dataset contains $\sim 180,000$ samples and and a 70/30 train/test split is applied.
The data consists of top quark jets with an energy of about 1~TeV, with each jet containing up to 30 particles.
These jet constituents are considered to be massless and can therefore be described by their 3-momenta or equivalently by transverse momentum $p_T$, pseudorapidity $\eta$, and azimuth angle $\phi$. In the \emph{JetNet} dataset, these variables are given relative to the jet momentum: 
$\eta_i^{rel} \coloneqq \eta_i^{particle}\!-\eta^{jet}$,
$\,\phi_i^{rel} \coloneqq (\phi_i^{particle}\!-\phi^{jet})\bmod 2\pi$, and 
$\,p_{T,i}^{rel} \coloneqq p_{T.i}^{particle}\!/p_T^{jet}$
, where $i$ runs over the particles in a jet.

The invariant mass $m_{jet}$ of a jet is an essential high-level feature containing important physics information. It is a global variable that depends on the correlations between the single jet constituents and provides therefore an important metric for the performance of the generative model. For the relative quantities above, we can define the scaled jet mass as $(m^{rel})^2=(\sum_i E^{~rel}_i)^2-(\sum_i \vec{p}^{~rel}_i)^2 = m_{jet}^2/p_{T,jet}^2$.


An additional set of metrics is constructed in~\cite{mpgan} by using the Energy Flow Polynomials (EFP)~\cite{efp}, which form a complete set of jet substructure observables.


The \emph{JetNet} paper additionally proposes a Message-Passing Generative Adversarial Network (MPGAN) that outperforms other state-of-the-art models at the time of publication and to which we compare our results.

\section{Generative Modelling}
Modeling an unknown complex probability function only from a set of data points is challenging. In addition, it is often not straightforward  to assess the quality of generated data. New learning paradigms are needed to be able to train such a model. GANs, as one of the most popular approaches, employ a setup of two networks, a generator and a discriminator that tries to distinguish generated and real data. Their training is notoriously difficult. A more stable approach to generative modelling are NFs, allowing Maximum Likelihood training, which is typically both fast and reproducible.
\subsection{Normalising Flow Models}
In NFs, the objective is to transform the training data distribution to a Normal distribution by mapping the training data with a series of invertible functions. 
This setup then allows for Maximum Likelihood optimization of the parameters of the transformation. Once the NF can transform the training data to a Normal distribution sufficiently well, new data is generated by sampling from a Normal distribution and applying the inverted transformations.  This study used the {\tt nflows}~\cite{nflows} library for NFs.

\subsubsection{Rational Quadratic Spline Coupling Layers}
In NFs with coupling layers, a clever trick is used to guarantee the invertibility of the transformation: the input dataset is split into two disjoint sets along its feature dimensions. The first set is mapped with the identity function, whereas the second set is transformed with a usually elementwise transformation. In the case of Rational Quadratic Splines (RQS)~\cite{nsf}, the quotient of two monotonic quadratic splines is used, which is by construction invertible. To increase the expressivity of the transformation, the parameters of the RQS for the second feature set are the output of a neural network applied to the first set. In our model, multiple of such RQS coupling layers are stacked, and after each layer a different split for the two sets is chosen. The Neural Networks used for producing the parameters of the RQS are fully connected feed-forward networks, using skip connections.
\subsection{Transformers}
Transformers were originally proposed as sequence-to-sequence models for machine translation~\cite{transformers}. However, they have been widely adopted in various fields like computer vision and speech processing.  On an abstract level, the main ingredients of a transformer are the encoder block and the decoder block. In this study, similarly as for most vision transformers, only an encoder block is used~\cite{BERT}. The attention architecture~\cite{attention} is typically permutation invariant, which is a useful feature for modelling particles in jets.
The encoder block consists of two main components, one being $1\times1$ convolutions applied token-wise (here particle-wise, since in this case a token corresponds to a particle) and the second being multi-headed self attention.
Important for our use case is that transformers can practically take a variable number of particles as input. This is done by zero-padding particles, which are then masked in the calculation of the attention: their respective values are set to negative infinity in the softmax function of the attention calculation such that they do not have any influence on the output at all.

In this study the output of the NFs is corrected with a post-processing transformer encoder network. An adversarial setup is used to train the refinement network which provides an additive correction to the output of the NF. A discriminator or critic is used to judge whether the corrected NF generated jets resemble real data. This refinement setup is used instead of a setup of only a generator and discriminator  as the training of adversarial architectures and transformers is already difficult enough without combining them.  
\begin{figure}
    \centering
    
     \includegraphics[width=1\textwidth]{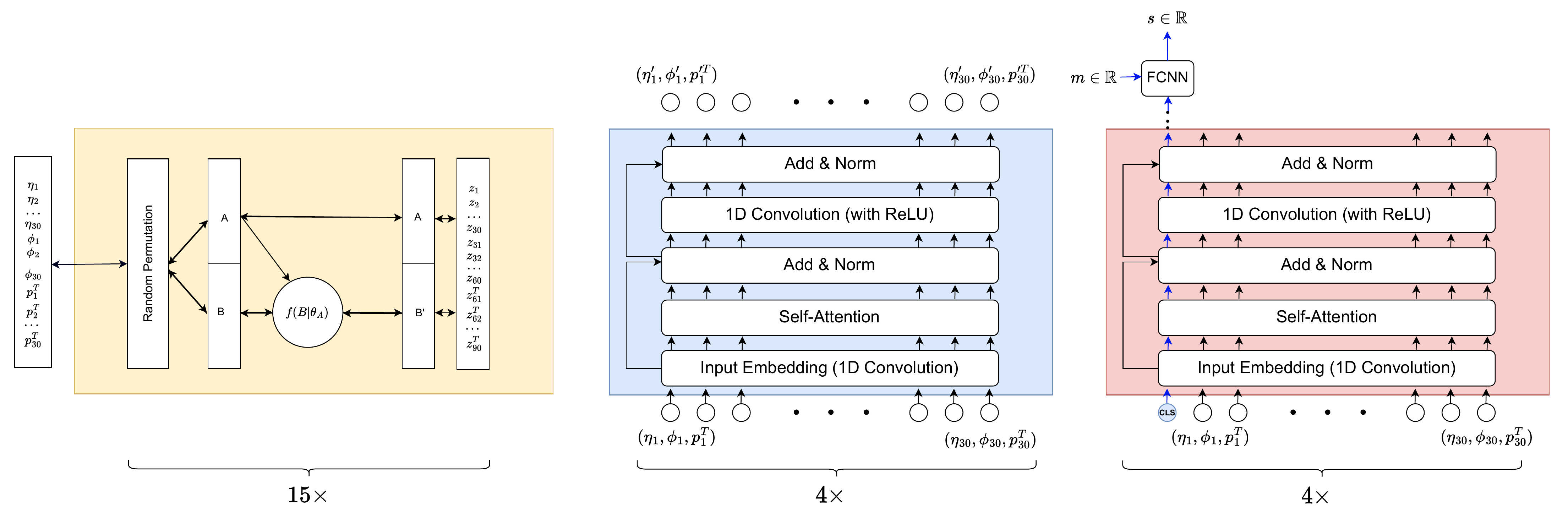}
     \caption{An overview of all components used in the model. The schematic on the left describes the NF, made up of 15 coupling layers. The middle one shows the refinement network, which is made up of 4 transformer encoder blocks and the one on the right similarly illustrates the critic network. All three models are used during the training, while only the first two are used during generation. The number of layers are given for the model architecture that gives the results presented in this study.} 
     \label{fig:architecture}
\end{figure}

\subsubsection{Post-processing network}
The post-processing network is a transformer encoder, which takes the output of the NF and reshapes it to $\mathbb{R}^{N\times D}$, where $N$ is the maximum number particles per jet and $D$ is the features per particle, in the case of this study $D=3$, and $N=30$. 
It uses Layer Normalization~\cite{Layernorm} and residual connections to connect the pre-attention features with the post-attention layer features (skip connections in fig. \ref{fig:architecture}). The normalization is applied after the addition (Add\&Norm in fig. \ref{fig:architecture}).

\subsubsection{Discriminator/Critic Network}
Depending on whether a more Vanilla GAN-like~\cite{GAN,LSGAN} or a WGAN~\cite{WGAN,wgangp} setup is employed, a discriminator or critic is used.
The latter is another transformer encoder that uses a BERT style classification token~\cite{BERT} to assess the realness of a jet. After the classification token passes through the second transformer encoder network, a fully-connected network is applied, which additionally takes the invariant mass as input. The rest of the architecture is the same as for the post-processing network. A schematic of all parts of the architecture is shown in Fig.~\ref{fig:architecture}.

\subsection{Training}
Different Training procedures are tested and compared, including the Vanilla GAN~\cite{GAN} training, an LSGAN~\cite{LSGAN} and a WGAN training with gradient penalty~\cite{WGAN,wgangp}. Optimization is done with RMSprop ~\cite{rmsprop}, the learning rate was varied with a two-cycle cosine scheduling after a linear warm-up that is commonly used with transformers. The most stable results were obtained with the LSGAN training. The presented model was trained on a NVIDIA P100 GPU after around 11 hours of training time. To produce results which can be compared to Ref.~\cite{mpgan}, we perform the same (70/30) train/test split  and evaluate with the functions that are provided by the \emph{JetNet} library.

\section{Results}
\begin{figure}
  \centering
  \includegraphics[width=\textwidth]{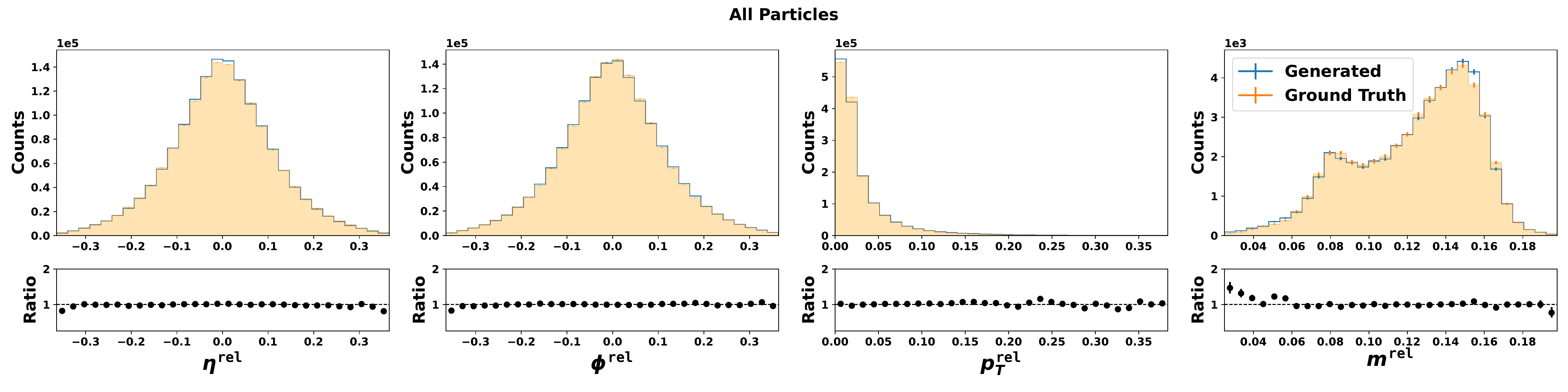}
  \caption{Distribution of all particles $(\eta^{rel},\phi^{rel},p_T^{rel})$ and the invariant mass. The generated data is compared to a hold out data set, and a ratio is shown below the plot.}
  \label{fig:results}
\end{figure}

The results on the top-quark dataset from Ref.~\cite{jetnet} are shown in Table~\ref{tab:results}. The first two entries are from the best performing models of Ref.~\cite{mpgan}, while VNF is the evaluation of the NF output and TF is the transformer-refined results. The best performing model used 15 RQS coupling layers and 4 transformer encoder layers in the refinement network and the critic. The  performance of the proposed models is comparable or better with respect to all considered metrics. Of particular interest is the FPND metric, where the proposed model performs considerably better.
In Fig.~\ref{fig:results} the results are visualized by comparing the generated particles with the holdout set. The marginal distribution of all features of all particles and the distribution of the invariant mass of the jet are shown. The model needs $\sim 7.6 \mu s$ for the generation of one jet, which is 4.6 times faster than the model in Ref.~\cite{mpgan} on a {\tt NVIDIA A100}.

\begin{table}[h]
\centering
\begin{tabular}{@{}l|llllll@{}}
\toprule
 Model & $W_1^M (\times 10^{-3})$ & $W_1^P
(\times 10^{-3})$ & $W_1^{EFP}(\times 10^{-5})$ &  FPND & COV $\uparrow$ &   MMD \\
\cline{1-7}



MP-MP~\cite{mpgan} & $\mathbf{0.6 \pm 0.2}$ &$2.3 \pm 0.3$ &$\mathbf{2 \pm 1}$ &$0.37$ & $\mathbf{0.57}$ & $\mathbf{0.071}$ \\
MP\_LFC-MP~\cite{mpgan} &$0.9 \pm 0.3$ &$2.2 \pm 0.7$ &$\mathbf{2 \pm 1}$ &$0.93$ &$0.56$ &$0.073$ \\
VNF &$5.0 \pm 0.2$ &$2.5 \pm 0.6$ &$14 \pm 2$ &$5.61$ &$0.56$ &$0.072$ \\
 TF &$0.78 \pm 0.09$ & $\mathbf{1.3 \pm 0.3}$ &$\mathbf{2 \pm 1}$ & $\mathbf{0.08}$ & $\mathbf{0.57}$ &$0.072$ \\\cline{1-7}
 \bottomrule
\end{tabular}
\vspace{20 pt}
\caption{Comparison between the best performing models from Ref.~\cite{mpgan}, with different models from this study on the top quark dataset. Scores written in bold highlight the best value.
}
\label{tab:results}

\end{table}

\section{Conclusion}
In this study a particle cloud generation model consisting of normalising flows and transformer encoders is proposed. The transformer encoder acts as a refinement model on the output of the normalising flow, and is trained in an adversarial fashion using another transformer encoder as critic. Intuitively one would not expect that NFs are well suited for point cloud generation, however they provide a  stable starting point, as adversarial setups and transformer models are notoriously difficult to optimize. The model reaches state-of-the-art performance on the publicly available \emph{JetNet}~\cite{jetnetlib} top quark dataset. 
It is important to note that particle clouds are similar to point clouds, possibly even more complex. As such it is expected that the model also would perform well on point cloud generation tasks.
For future studies it is planned to use the model on the other \emph{JetNet} datasets~\cite{jetnet} as well as on a dataset with more particles, and to evaluate its scaleability. It is expected that inputs with a higher dimensionality are difficult for the attention calculation as it scales with the number particles squared. But as transformers are an active field of research, there seem to be possible ways to overcome this~\cite{googleattention,otherattentionthing}.

\section{Acknowledgements}

Benno K\"ach is funded by Helmholtz Association’s Initiative and Networking Fund through Helmholtz AI (grant number: ZT-I-PF-5-64).
This research was supported in part through the Maxwell computational resources operated at Deutsches Elektronen-Synchrotron DESY (Hamburg, Germany).
The authors acknowledge support from Deutsches Elektronen-Synchrotron DESY (Hamburg, Germany), a member of the Helmholtz Association HGF.


\newpage
\newpage
\bibliographystyle{unsrtnat}


\bibliography{Untitled}  

\newpage
\section*{Checklist}



\begin{enumerate}

\item For all authors...
\begin{enumerate}
  \item Do the main claims made in the abstract and introduction accurately reflect the paper's contributions and scope?
    \answerYes{}
  \item Did you describe the limitations of your work?
    \answerYes{} Scalability is unclear
  \item Did you discuss any potential negative societal impacts of your work?
    \answerNo{} The model brings the same negative impact as most generative models -  but due to the limited space in this extended abstract it has been left out.
  \item Have you read the ethics review guidelines and ensured that your paper conforms to them?
   \answerYes{}
\end{enumerate}

\item If you are including theoretical results...
\begin{enumerate}
  \item Did you state the full set of assumptions of all theoretical results?
    \answerNA{}
        \item Did you include complete proofs of all theoretical results?
    \answerNA{}
\end{enumerate}

\item If you ran experiments...
\begin{enumerate}
  \item Did you include the code, data, and instructions needed to reproduce the main experimental results (either in the supplemental material or as a URL)?
    \answerNo{}
  \item Did you specify all the training details (e.g., data splits, hyperparameters, how they were chosen)?
    \answerNo{} Again, left out due to limited space
        \item Did you report error bars (e.g., with respect to the random seed after running experiments multiple times)?
    \answerNo{} Again, left out due to limited space
        \item Did you include the total amount of compute and the type of resources used (e.g., type of GPUs, internal cluster, or cloud provider)?
    \answerYes{}
\end{enumerate}

\item If you are using existing assets (e.g., code, data, models) or curating/releasing new assets...
\begin{enumerate}
  \item If your work uses existing assets, did you cite the creators?
    \answerYes{} The libraries that were used are \textit{nflows} and \textit{JetNet}
  \item Did you mention the license of the assets?
    \answerYes{} Both libraries use the MIT license
  \item Did you include any new assets either in the supplemental material or as a URL?
    \answerNo{}
  \item Did you discuss whether and how consent was obtained from people whose data you're using/curating?
    \answerNA{}
  \item Did you discuss whether the data you are using/curating contains personally identifiable information or offensive content?
    \answerNA{}
\end{enumerate}

\item If you used crowdsourcing or conducted research with human subjects...
\begin{enumerate}
  \item Did you include the full text of instructions given to participants and screenshots, if applicable?
    \answerNA{}
  \item Did you describe any potential participant risks, with links to Institutional Review Board (IRB) approvals, if applicable?
    \answerNA{}
  \item Did you include the estimated hourly wage paid to participants and the total amount spent on participant compensation?
    \answerNA{}
\end{enumerate}

\end{enumerate}


\end{document}